# Microcalorimetric and SAXS Determination of PEO-SDS Interactions: The Effect of Cosolutes Formed by Ions

**Aparecida Mageste Barbosa,[†] Igor José Boggione Santos,[†] Guilherme Max Dias Ferreira,[†] Maria do Carmo Hespanhol da Silva,[†] Alvaro Vianna Novaes de Carvalho Teixeira,[‡] and Luis Henrique Mendes da Silva*,[†]**

[†]*Grupo de Química Verde Coloidal e Macromolecular, Departamento de Química and*
[‡]*Departamento de Física, Centro de Ciências Exatas e Tecnológicas, Universidade Federal de Viçosa, Avenida P. H. Rolfs, Viçosa, MG, 36570-900, Brazil.*

The effect of different ionic cosolutes (NaCl, $Na_2SO_4$, $Li_2SO_4$, NaSCN, $Na_2[Fe(CN)_5NO]$, and $Na_3[Co(NO)_6]$) on the interaction between sodium dodecyl sulfate (SDS) and poly(ethylene oxide) (PEO) was examined by small-angle X-ray scattering (SAXS) and isothermal titration calorimetric techniques. The critical aggregation concentration values (cac), the saturation concentration ($C_2$), the integral enthalpy change for aggregate formation ($\Delta H_{agg}(int)$) and the standard free energy change of micelle adsorption on the macromolecule chain ($\Delta\Delta G_{agg}$) were derived from the calorimetric titration curves. In the presence of 1.00 mmol $L^{-1}$ cosolute, no changes in the parameters were observed when compared with those obtained for SDS-PEO interactions in pure water. For NaCl, $Na_2SO_4$, $Li_2SO_4$, and NaSCN at 10.0 and 100 mmol $L^{-1}$, the cosolute presence lowered cac, increased $C_2$, and the PEO-SDS aggregate became more stable. In the presence of $Na_2[Fe(CN)_5NO]$, the calorimetric titration curves changed drastically, showing a possible reduction in the PEO-SDS degree of interaction, possibility disrupting the formed nanostructure; however, the SAXS data confirmed, independent of the small energy observed, the presence of aggregates adsorbed on the polymer chain.

## 1. Introduction

There is general scientific agreement that complex fluids formed by mixtures of aqueous solutions of surfactants and macromolecules exhibit intriguing properties that are highly valued in formulations of paints, coatings, cosmetic products, agrochemicals, and laundry detergents.[1] It is now recognized that these new mixture properties arise from a balance of relatively weak binding forces (hydrophobic, dipole-dipole, ionic-dipole and dispersion),[2] which control the thermodynamic process of structure formation between the surfactant and polymer. At low surfactant concentrations (or small surfactant-to-polymer ratios, $R_{s/p}$), individual molecules adsorb along the polymer, which is characterized by a critical aggregation concentration, cac. At intermediate $R_{s/p}$ values, surfactant monomers aggregate close to the macromolecule chain. After polymer molecule saturation (saturation concentration, $C_2$), further addition of surfactant (increase of $R_{s/p}$) promotes micelle formation in pure water.[3] Previous studies have corroborated the molecular processes described above, including surface tension measurements,[4] conductivity,[5] dialysis,[6] viscosity,[7] dye solubilization,[8] microcalorimetry,[9] and scattering techniques.[10] The topic has also been treated in several very good review articles[11] and book chapters.[12] However, most of the studies that have been carried out deal with polymer-surfactant interactions in pure aqueous microenvironments, but there are few studies of interactions occurring in electrolyte aqueous solutions.[13] Almost all studies have investigated the effect of simple inorganic salts, NaCl or NaBr, with the exception of Saito's works, which investigated the effect of large organic ions.[14] Electrolytes in the presence of polymer-surfactant complexes generally decrease the critical aggregation concentration (cac) and increase the binding ratio of surfactant to polymer ($C_2$). Dubin et al.[15] suggested that surfactant counterions or cations from added salts play a role by simultaneously interacting with micelles and PEO, leading to an increase in the number of micelles bound per chain with increasing ionic strength. To the best of our knowledge, all investigations on anionic effects on polymer-surfactant interactions suggest no or only small contributions to surfactant adsorption on the polymer chain. In 2006, da Silva et al.[16] proposed an interaction between nitroprusside anion, $[Fe(CN)5NO]^{2-}$, and poly(ethylene oxide). On the basis of their results, it is reasonable to propose that there should be a specific enthalpic interaction between the ion and the macromolecule in order for the complex anion to concentrate in the polymer-rich phase of an aqueous two-phase system (ATPS).[17,18] The enthalpic interaction between nitroprusside ion and poly(ethylene oxide) macromolecules probably occurs between the $[Fe(CN)_5NO]^{2-}$ and ethylene oxide units and is likely very dependent on the nature of the central atom in the complex $[M(CN)_5NO]^{x-}$ ($M$ = Fe, Mn, and Cr).[19] Recently, we discovered that the anion $[Co(NO_2)_6]^{2-}$ concentrated in the top phase of ATPSs, indicating a possible specific EO-$[Co(NO_2)_6]^{2-}$ interaction.

Our aim in the present work was to identify ionic cosolute effects on the driving forces associated with polymer-surfactant interactions, mainly relating to anion contributions. The effects of the electrolytes NaCl, $Na_2SO_4$, $Li_2SO_4$, NaSCN, $Na_2[Fe(CN)_5NO]$, and $Na_2[Co(NO_2)_6]$ on the micellization and binding interaction between SDS and PEO was examined by isothermal titration calorimetric and small-angle X-ray scattering (SAXS) techniques.

* To whom correspondence should be addressed: Phone: +55 31 38993052. Fax: +55 31 38993065. E-mail: luhen@ufv.br.
† Grupo de Química Verde Coloidal e Macromolecular, Departamento de Química.
‡ Departamento de Física.

## 2. Experimental Section

**2.1. Materials.** Poly(ethylene oxide) with an average molar mass (according to the manufacturer) of 35 000 g mol$^{-1}$ (designated as PEO35k) was supplied by Fluka (U.S.A.). Sodium dodecyl sulfate (SDS), purchased from Fluka, was of the highest purity available (g99.0%). The critical micelle concentrations (cmc) values for the surfactant were in agreement with data in the literature.[20] Cosolutes such as Na$_2$SO$_4$, Li$_2$SO$_4$, NaSCN, and Na$_2$[Co(NO$_2$)$_6$] were manufactured by Vetec (Brazil), while Na$_2$[Fe(CN)$_5$NO] and NaCl were from Merck (U.S.A.). All chemicals were used without further purification.

**2.2. Isothermal Titration Calorimetry.** The enthalpy changes of PEO and surfactant interactions in the presence of four different concentrations of ionic cosolutes (0.0, 1.00, 10.0, and 100 mmol L$^{-1}$) were performed in triplicate using a CSC-4200 microcalorimeter (Calorimeter Science Corp.) controlled by ItcRun software with a 1.75 mL reaction cell (sample and reference). The whole calorimetric procedure was chemically and electrically calibrated to the heat of protonation of (tris(hydroxymethyl) aminomethane) and the joule effect, as recommended.[21] Each cosolute aqueous solution was used as a solvent in the preparation of 0.100 wt % PEO and 10.0 wt % SDS solutions. Deionized water was used for preparing all solutions. The titrations were carried out by step-by-step injections (5 μL) of a concentrated surfactant titrant solution with a gastight Hamilton syringe (250 μL), controlled by an instrument, with intervals of 10 min between each injection. Aliquots of concentrated surfactant solution, dissolved in pure water or in an aqueous solution of cosolute, were added to a sample cell containing pure aqueous solution of PEO or a mixture formed by dissolving the polymer in an aqueous solution of a cosolute. The solution was titrated in the sample cell with stirring at 300 rpm using a helix stirrer, and measurements were carried out at a constant temperature of 25.000 ± 0.001 °C.

**2.3. Small-Angle X-Ray Scattering.** Small angle scattering measurements were performed at the D02A SAXS2 beamline in the Brazilian Synchrotron Light Laboratory (LNLS, Campinas-SP).

Na$_2$[Fe(CN)$_5$NO] and Na$_3$[Co(NO)$_6$]) solutions were prepared at three different concentrations, 1.00, 10.0, and 100 mM. These solutions were used as solvents in the preparation of solutions of 0.100 wt % PEO and 10.0 wt % SDS. Samples of 10.0 mL flasks were charged with 75, 125, 175, 250, 350, and 775 μL of SDS and were made up to volume with a solution of PEO.

The data were collected using a CCD (MAR Research) at two sample-detector distances, 2503 and 383 mm, and the X-ray wavelength was fixed at $\lambda = 0.1488$ nm. In this way, the covered values of the scattering vector $q = 4\pi/\lambda \sin(\theta/2)$, where $\theta$ is the scattering angle, were from 0.07 to 9 nm$^{-1}$. Data reduction to $I(q)$ vs $q$ and solvent subtractions were performed with the program FIT2D. Standard corrections and error calculations were included in this routine.

## 3. Results and Discussion

By using isothermal titration calorimetry to investigate polymer-surfactant interactions, it is possible to obtain five important parameters that characterize these interactions, namely (i) the critical aggregation concentration value, (ii) the saturation concentration, (iii) the integral enthalpy change for aggregate formation, $\Delta H_{agg}$ (int), (iv) the standard free energy of micelle adsorption on the macromolecule chain $\Delta\Delta G_{agg}$, and (v) the number of moles of bound surfactant per PEO unit in the polymer. Additionally, based on features of the apparent enthalpy curves, one can obtain qualitative information about the progress of aggregation with increasing SDS concentration.[22-26] Figure 1 shows the titration curves obtained in our work, where the observed enthalpy changes, $\Delta H_{obs}$, for each injection are plotted against the total SDS concentration. In a typical experiment, there was an addition of 5 μL of an SDS aqueous solution (10.0 wt %) to (i) the dilute PEO aqueous solution (0.100 wt %) and to (ii) pure water. Our results were in agreement with those of other groups[27-29] and showed a very small enthalpy change associated with the SDS micellization process, which only caused a change in the dilution curve slope at the cmc (8.3 mmol L$^{-1}$). At the start of the titration experiment, both curves (in PEO solutions and in pure water) were coincident, indicating that at very small surfactant concentrations, there is no calorimetrically detectable interaction between PEO and SDS. However, at cac = 3.6 ± 0.1 mmol L$^{-1}$ of SDS, the titration curve in the PEO solution started to deviate from the SDS dilution curve, showing a pronounced endothermic peak followed by a broad shallow exothermic one (relative to the dilution curve in water). This onset of surfactant aggregation in the presence of polymer was smaller than the cmc, which is an indication that polymer-SDS interactions make the SDS micelles adsorbed at the polymer interface more stable than similar aggregates dissolved in bulk solution. By making assumptions that the driving force for surfactants aggregating onto polymers is similar to that for normal free surfactant micellization processes, the standard free energy of micelle adsorption on the macromolecule chain can be estimated from $\Delta\Delta G_{agg} = RT \ln(cac/cmc)$. For the PEO-SDS interaction in pure water, $\Delta\Delta G_{agg} = -2.08 \pm 0.03$ kJ mol$^{-1}$. This stabilization mainly arises from the solubilization of EO groups in the headgroup region of the micelles with a concomitant decrease in electrostatic repulsion. At a total SDS concentration equal to 17.5 ± 1.1 mmol L$^{-1}$, the PEO chain became saturated with SDS molecules, the influence of the polymer on the aggregation of surfactant in the calorimeter cell ceased, the free monomer concentration reached the cmc and free micelles started to form. This critical concentration, defined as the concentration where the titration curve in polymer solution joins the dilution curve in water, is known as $C_2$.

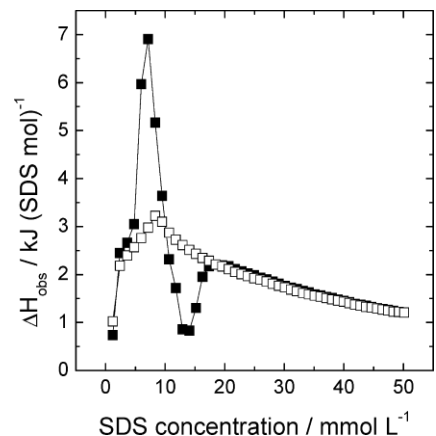

**Figure 1.** Calorimetric titration curves from (■) the addition of 10.00 wt % SDS aqueous solution to 0.10 wt % PEO 35 000 aqueous solution and (□) dilution in water at 25 °C.

In order to evaluate the differential enthalpy change for the PEO-SDS interactions, we must subtract the titration curve in polymer solution from the dilution curve in water at each SDS concentration (Figure 2).

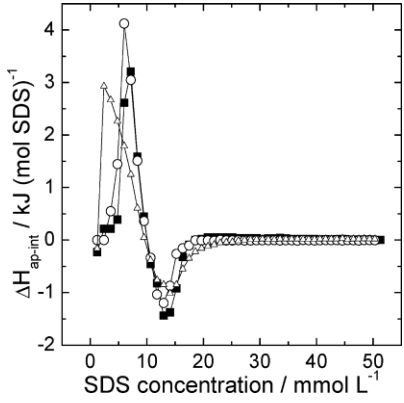

**Figure 2.** Apparent molar enthalpy change of interaction between PEO35000 and SDS in (■) pure water, (○) 1.0 mmol L$^{-1}$ of NaCl, and (△) 100 mmol L$^{-1}$ of NaCl at 25 °C.

Unfortunately, as the extent of binding (amount of aggregates formed) is not known, we cannot calculate the exact molar enthalpy change of interaction, only an apparent molar enthalpy change, $\Delta H_{ap\text{-}int}$. However, the features of the $\Delta H_{ap\text{-}int}$ curve give qualitative information about the progress of aggregation with increasing SDS concentration.[30-32] It is well established that the endothermic peak is associated with the SDS and PEO segments' dehydration processes, while the exothermic peak, observed at higher SDS concentrations, is attributed to rehydration of the previously dehydrated PEO segments.[33]

In the absence of isothermal binding data of SDS to PEO, we can calculate the integral enthalpy change for aggregate formation, $\Delta H_{agg}(int)$, which expresses the enthalpy change of the formation of one mole of aggregated surfactant from monomers over the concentration range from cac up to $C_2$. Following Olofsson and Loh,[34] for the $\Delta H_{agg}(int)$ calculation, we assumed that $C_2$ had been reached after $Y$ injections of concentrated surfactant solution (each injection added $n^{inj}$ mols of surfactant) to give a total volume of $V_Y$. At $C_2$, we added a total surfactant equal to $[Yn^{inj}]$, but from this total, $[V_Y(cmc)]$ moles do not interact with the PEO segments. Naturally, the total energy measured, $[\Sigma q_{obs}]$, should be discounted by the energy of demicellization and dilution, $[Yq_{demic+dil}]$. Mathematically, $\Delta H_{agg}(int)$ is calculated by eq 1

$$\Delta H_{agg}(int) = \left[\frac{\Sigma q_{obs} - \Sigma q_{demic+dil}}{Yn^{inj} - V_Y(cmc)}\right] \quad (1)$$

A summary of all of these SDS-PEO parameters, in the presence or absence of ionic cosolutes, is presented in Table 1.

### 3.1. Effects of Simple Salts on the PEO-SDS Interaction.

The presence of different electrolytes, which can effectively modulate the solvent quality of water, could have a significant influence on the PEO-SDS aggregation characteristics. The water-structure forming salts such as NaCl, Na$_2$SO$_4$ and Li$_2$SO$_4$ decrease the solubility of both polymer and surfactant (salting out effect) and, probably, could reduce their critical aggregation parameters.

Columns 2 and 3 in Table 1 show the effect of NaCl, Na$_2$SO$_4$, Li$_2$SO$_4$, and NaSCN on the critical aggregation concentration (cac) and saturation concentration ($C_2$) at three different cosolute concentrations (1.00, 10.0, and 100 mmol L$^{-1}$). In general, the magnitude of the effect was dependent on the electrolyte nature, but at 1.00 mmol L$^{-1}$, no changes in either parameter were noted when compared with the values measured in pure water. However, for other cosolute concentrations an increase in the salt concentration promoted an increase in $C_2$ (and consequently on the extent of binding), which means that higher ionic forces make it possible for more surfactant monomers to be adsorbed on the polymer chain. In addition, the presence of more electrolytes caused a decrease in the cac values, indicating that the association starts at lower surfactant concentrations. This is a general behavior that has been observed in other studies.[8,15,35-37] According to the fundamental Gibbs equation of adsorption, $\Gamma = -(d\gamma/d\mu)$, both salt effects could be attributed to a decrease in the interfacial tension between the PEO surface and water, $d\gamma < 0$, and/or an increase in the surfactant chemical potential, $d\mu_s > 0$. The only manner through which a salt could decrease $\gamma$ is by specific interactions between the ions and the PEO segments. The adsorption and distribution of ions at interfaces of macromolecules/water is a fundamental process encountered in a wide range of biological and chemical systems.[38] With PEO, it is recognized that the main interaction of PEO is with cations.[39] However, Quina et al.[40] followed the fluorescence quenching of free and polymer-bound chromophores by several salts (NaI, LiI, KI, NaSCN, LiSCN, and KSCN) in water and methanol. They observed that with iodide and thiocyanate anions, quenching was only observed with the polymer, demonstrating that the PEO-ion interaction exists with the following order of quenching: $Li^+ < Na^+ < K^+ < Cs^+ < Rb^+$. Moreover, when these experiments were repeated with NaCl and KCl, no quenching was observed, indicating that it was due to the anion and, hence, revealing an increased local concentration of the anion in the vicinity of the polymer. Therefore, any interaction between electrolytes and PEO must involve both cations and anions. Recently, da Silva and Loh[41] attributed the trend in efficacy of sodium and lithium sulfates at inducing ATPS formation to cation-polymer interactions based on calorimetric measurements. In the same work, the authors demonstrated that NaCl did not interact with PEO. Their proposed model suggested that when PEO and sulfate salts are mixed, the cations and the polymer interact, releasing some water molecules that were solvating them, which is driven by the entropy increase. This cation binding continues as more electrolytes are added until a saturation point is reached, after which no more entropy gain may be attained and phase splitting becomes more favorable. Therefore, the picture that arises from this proposed model is that in systems containing macromolecules bound to cations, there is the formation of a pseudopolycation, which is capable of interacting with negatively charged species. In fact, Dubin et al.[15] suggested that surfactant counterions play a role by simultaneously interacting with micelles (through electrostatic forces) and PEO, and that the number of micelles bound per chain increases with ionic strength. On the basis of our results and the above discussion, it is evident that the NaCl effect, a salt that does not interact with PEO, is due to an increase in the surfactant chemical potential, leading to a decrease not only in the cac, but also in the cmc. As the magnitude of the effects caused by others salts is the same as that of NaCl, and because it was demonstrated that they interact with macromolecules, we suggest that the possible interactions between cosolute ions and PEO segments are not intense enough to make a significant change in the thermodynamics of the PEO-SDS interaction.

**TABLE 1: Thermodynamics Parameter for PEO-SDS Interactions in the Presence of Simple Salts**[a]

| PEO35000/SDS system in water or salt | cac/mmol L$^{-1}$ | $C_2$/mmol L$^{-1}$ | cmc/mmol L$^{-1}$ | $\Delta H_{agg}(int)$/kJ mol$^{-1}$ | $\Delta\Delta G_{agg}$/kJ mol$^{-1}$ | extent of binding/(mmol g$^{-1}$) of polymer |
|---|---|---|---|---|---|---|
| pure H$_2$O | 3.6±0.1 | 18.6±1.1 | 8.3±0.1 | -1.05±0.04 | -2.1±0.0 | 19.4±2.0 |
| NaCl, 1.00 mmol L$^1$ | 3.6±0.1 | 18.6±1.1 | 8.3±0.1 | +0.64±0.02 | -2.1±0.0 | 19.4±1.8 |
| NaCl, 100 mmol L$^1$ | 2.4±0.2 | 26.2±1.7 | 5.9±0.1 | +0.35±0.01 | -2.2±0.1 | 27.9±2.3 |
| Na$_2$SO$_4$, 1.00 mmol L$^1$ | 3.6±0.2 | 18.6±1.1 | 8.3±0.1 | -0.63±0.04 | -2.1±0.1 | 19.4±1.6 |
| Na$_2$SO$_4$, 10.0 mmol L$^1$ | 2.4±0.1 | 25.1±1.4 | 7.5±0.1 | +0.07±0.01 | -2.8±0.0 | 26.7±1.8 |
| Na$_2$SO$_4$, 100 mmol L$^1$ | 2.4±0.1 | 27.3±1.4 | 4.0±0.2 | +0.03±0.01 | -1.3±0.1 | 29.1±2.1 |
| Li$_2$SO$_4$, 1.00 mmol L$^1$ | 3.6±0.1 | 18.6±1.1 | 8.3±0.1 | -0.63±0.04 | -2.1±0.0 | 19.4±1.0 |
| Li$_2$SO$_4$, 10.0 mmol L$^1$ | 2.4±0.1 | 19.7±1.1 | 6.0±0.3 | -1.00±0.04 | -2.3±0.1 | 20.6±1.2 |
| Li$_2$SO$_4$, 100 mmol L$^1$ | 2.4±0.1 | 25.1±1.1 | 4.8±0.2 | -0.40±0.03 | -1.7±0.1 | 26.7±1.6 |
| NaSCN, 1.00 mmol L$^1$ | 3.6±0.1 | 18.6±1.1 | 7.2±0.2 | +1.84±0.04 | -1.7±0.0 | 19.4±0.8 |
| NaSCN, 10.0 mmol L$^1$ | 3.6±0.1 | 20.8±1.1 | 4.8±0.1 | -0.76±0.02 | -0.7±0.0 | 23.1±1.1 |
| NaSCN, 100 mmol L$^1$ | 2.4±0.1 | 24.1±1.1 | 4.6±0.1 | +0.23±0.01 | -1.6±0.1 | 25.5±1.7 |

[a] cac, critical aggregation concentration; $C_2$, saturation concentration; cmc, critical micelle concentration; $\Delta H_{agg}(int)$, integral enthalpy change for aggregate formation; $\Delta\Delta G_{agg}$, standard free energy of micelle adsorption.

The change in the free energy of aggregation, $\Delta\Delta G_{agg}$, is a quantitative measure of how much more stable the surfactant aggregate formed in the presence of polymer is when compared with the normal, free surfactant micellization process. Interestingly, the electrolytes were capable of increasing the amount of adsorbed surfactant (higher extent of binding) without significantly changing $\Delta\Delta G_{agg}$, which means that ion-ion interactions are not a significant contribution to the delicate balance of forces responsible for the PEO-SDS interaction. This behavior corroborates the point of view that the motriz power of the SDS-PEO interaction is hydrophobic in nature.[42] The small decrease in $\Delta\Delta G_{agg}$ values observed when more surfactant is incorporated on the polymer chain (due to the presence of electrolyte) could be attributed to electrostatic repulsion between the surfactant's hydrophilic region; the electrostatic repulsion is not very intense because salt stabilizes surfactant aggregates by effectively screening electrostatic interactions in the micellar surface.[15] The salt effect was pronounced on the $\Delta H_{agg}(int)$ parameter and the $\Delta H_{ap-int}(dif)$ curves (Figures 2-5).

For all salts, except $Li_2SO_4$, the interactions promoted by the increase in ion concentration in the system were capable of changing the aggregation process from exothermic to endothermic, that is, $\Delta H_{agg}(int) > 0$, again highlighting the hydrophobic character of the PEO-SDS interaction. On the basis of the $\Delta H_{ap-int}(dif)$ curves, it is evident that the salt effect is mainly on the solvation shell of the interacting particles, since the peaks of both curves, endothermic and exothermic, decreased with an increase in the salt concentration. For $Li_2SO_4$, $\Delta H_{agg}(int)$ was exothermic at all concentrations, suggesting that this electrolyte interacts with EO segments, reducing the degree of hydration of the polymer and consequently decreasing the energy necessary

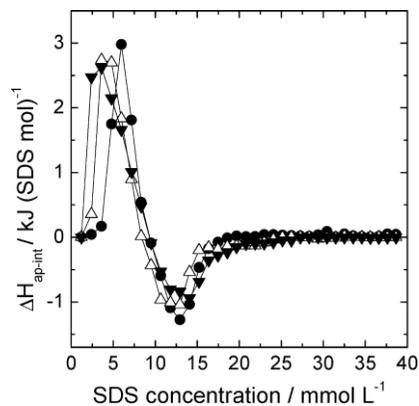

**Figure 3.** Apparent molar enthalpy change of interaction between PEO35000 and SDS in $Na_2SO_4$ aqueous solutions: (●) 1.0 mmol L$^{-1}$, (△) 10.0 mmol L$^{-1}$, and (▼) 100.0 mmol L$^{-1}$ at 25 °C.

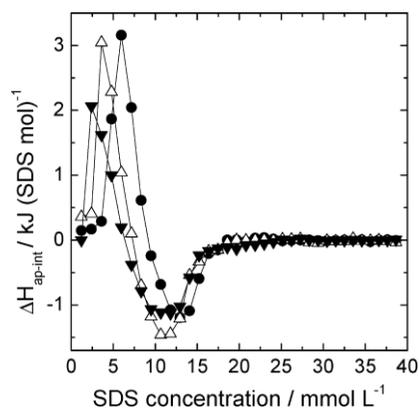

**Figure 4.** Apparent molar enthalpy change of interaction between PEO35000 and SDS in $Li_2SO_4$ aqueous solutions: (●) 1.0 mmol L$^{-1}$, (△) 10.0 mmol L$^{-1}$, and (▼) 100.0 mmol L$^{-1}$ at 25 °C.

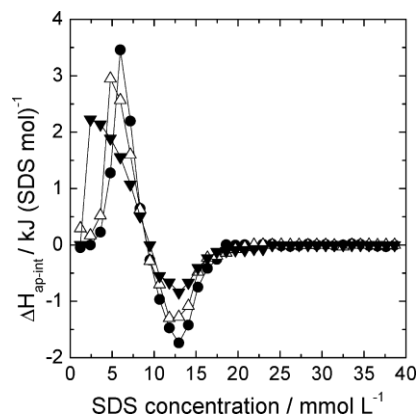

**Figure 5.** Apparent molar enthalpy change of interaction between PEO35000 and SDS in NaSCN aqueous solutions: (●) 1.0 mmol L$^{-1}$, (△) 10.0 mmol L$^{-1}$, and (▼) 100.0 mmol L$^{-1}$ at 25 °C.

for dehydration of the PEO segments (see the endothermic peak of 100 mmol L$^{-1}$).[15]

### 3.2. Effects of Complex Salts on the PEO-SDS Interaction.
Recent research has shown that anions play an important role in determining the self-assembly behavior of some kinds of surfactants, including some macromolecular amphiphilic agents.[43] There have been various theoretical and experimental efforts aimed at explaining these effects; however, no definitive model has been put forth. Some experimental results indicate that the anion and the polymer strongly interact, similar to complex formation, and that this complex presents a lower aqueous solubility.

**TABLE 2: Thermodynamics Parameter for PEO-SDS Interactions in the Presence of Complex Salts**[a]

| PEO35000/SDS system in water or salt | cac/mmol L$^{-1}$ | $C_2$/mmol L$^{-1}$ | cmc/mmol L$^{-1}$ | $\Delta H_{agg}$(int)/kJ mol$^{-1}$ | $\Delta\Delta G_{agg}$/kJ mol$^{-1}$ | extent of binding/ mmol/g of polymer |
|---|---|---|---|---|---|---|
| pure H$_2$O | 3.6 ± 0.1 | 18.6 ± 1.1 | 8.3 ± 0.1 | −0.91 ± 0.04 | −2.1 ± 0.0 | 19.4 ± 2.0 |
| Na$_2$[Fe(CN)$_5$NO], 1.00 mmol | 3.6 ± 0.1 | 21.9 ± 1.1 | 7.2 ± 0.1 | −0.82 ± 0.05 | −1.7 ± 0.0 | 23.1 ± 1.1 |
| Na$_2$[Fe(CN)$_5$NO], 10.0 mmol | 3.6 ± 0.1 | 23.0 ± 1.1 | 6.0 ± 0.1 | +0.03 ± 0.01 | −1.3 ± 0.0 | 24.3 ± 0.8 |
| Na$_2$[Fe(CN)$_5$NO], 100 mmol L$^{-1}$ | 2.4 ± 0.2 | 31.5 ± 1.1 | 3.6 ± 0.1 | −2.97 ± 0.02 | −1.0 ± 0.0 | 34.0 ± 1.0 |
| Na$_3$[Co(NO$_2$)$_6$], 1.00 mmol L$^{-1}$ | 2.4 ± 0.2 | 20.8 ± 1.1 | 8.3 ± 0.1 | −0.17 ± 0.03 | −3.1 ± 0.1 | 21.9 ± 0.9 |
| Na$_3$[Co(NO$_2$)$_6$], 10.0 mmol L$^{-1}$ | 2.4 ± 0.1 | 23.0 ± 1.4 | 6.0 ± 0.1 | +0.15 ± 0.01 | −2.3 ± 0.0 | 24.3 ± 1.0 |
| Na$_3$[Co(NO$_2$)$_6$], 100 mmol L$^{-1}$ | 2.4 ± 0.1 | 26.2 ± 1.4 | 6.0 ± 0.2 | −0.11 ± 0.02 | −2.3 ± 0.0 | 27.9 ± 1.0 |

[a] cac, critical aggregation concentration; $C_2$, saturation concentration; cmc, critical micelle concentration; $\Delta H_{agg}$(int), integral enthalpy change for aggregate formation; $\Delta\Delta G_{agg}$, standard free energy of micelle adsorption.

In 2006, da Silva et al.[16] proposed the existence of a direct interaction between [Fe(CN)$_5$NO]$^{2-}$ anion and PEO segments, leading a favorable enthalpy of transfer from the bottom phase to the top phase of an aqueous two-phase system (ATPS). According to the authors, this specific interaction occurs on the NO site, and this proposition was supported by infrared spectroscopy measurements of [Fe(CN)$_5$NO]$^{2-}$ dissolved in water and in aqueous solutions of PEO, where it was possible to see that the NO stretching band was very sensitive to the PEO concentration, while the other absorptions remained constant. When sodium nitroprusside was dissolved in pure water, the NO wavenumber was observed at 1936 cm$^{-1}$. However, this value decreased when the PEO concentration was increased, reaching a limiting value of 1898 cm$^{-1}$ in pure liquid PEO. There was no dependence of the NO stretching frequency on the size of PEO, indicating a site-specific interaction caused by an increased local concentration of the anion in the vicinity of the polymer. This increased local anion concentration must be promoted by the adsorption of the cation to the PEO chain. The NO frequency shift could be explained considering the diamagnetic character of the Fe$^{II}$NO$^+$ species (usually described as low-spin) and its preferential solvation. Thus, in PEO aqueous solutions, water molecules, and EO segments will form solvation shells around anions with a radial distribution that will depend on the polymer concentration. The acceptor-donor interaction between the ion species (mainly at the NO$^+$ site) and its solvation molecules (due to the electron lone pairs present on the oxygen atom) will change the NO electron density, altering the force constants of the NO bond. More recently, our group discovered that the anion [Co(NO$_2$)$_6$]$^{3-}$ also interacts with PEO segments, probably due to the -NO$_2$ groups, based on partitioning behavior in an ATPS.[44] On the basis of the above discussion, we expected that the presence of these anions would change the interactions between PEO and SDS. Table 2 shows critical aggregation concentrations, saturation concentrations, critical micelle concentrations, integral enthalpy changes of aggregate formation, standard free energies of micelle adsorption, and extents of binding of surfactant for PEO-SDS interactions in the presence of the complex salts Na$_2$[Fe(CN)$_5$NO] and Na$_3$[Co(NO$_2$)$_6$]. As for simple salts, an increase in complex electrolyte concentration reduced the cac and increased the $C_2$. The extent of binding became higher with an increase in the complex salt concentration but without a significant increase in the $\Delta\Delta G_{agg}$. In fact, in the case of Na$_3$[Co(NO$_2$)$_6$], an increase in the amount of surfactant adsorbed on the polymer chain occurred with a decrease in the standard free energy of micelle adsorption, $\Delta\Delta G_{agg}$. On the basis of these parameters, it is possible to suggest that this kind of complex electrolyte does not interact with PEO segments. However, when we look at the curve of apparent enthalpy change of interaction, $\Delta H_{ap-int}$(dif), it is evident that there was a pronounced effect in the system containing the Na$_2$[Fe(CN)$_5$NO] (Figure 6) and a lower effect in the mixture PEO + SDS + Na$_3$[Co(NO$_2$)$_6$] (Figure 6). The Na$_3$[Co(NO$_2$)$_6$]-PEO interaction is so strong that at a lower concentration (1.00 mmol L$^{-1}$), this complex salt produces a change on the $\Delta H_{ap-int}$(dif) × [SDS] curve.

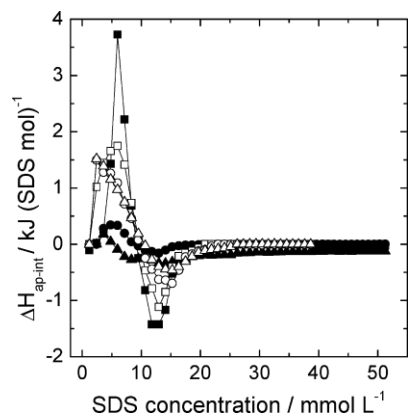

**Figure 6.** Apparent molar enthalpy change of interaction between PEO35000 and SDS in salt aqueous solutions. Na$_2$[Fe(CN)$_5$NO]: (■) 1.0 mmol L$^{-1}$, (●) 10.0 mmol L$^{-1}$, and (▲) 100.0 mmol L$^{-1}$ at 25 °C. Na$_3$[Co(NO$_2$)$_6$]: (□) 1.0 mmol L$^{-1}$, (○) 10.0 mmol L$^{-1}$, and (△) 100.0 mmol L$^{-1}$ at 25 °C.

For Na$_2$[Fe(CN)$_5$NO] at concentrations of 10.0 mmol L$^{-1}$ and 100 mmol L$^{-1}$, the $\Delta H_{ap-int}$(dif) was reduced to almost zero.

To analyze the spatial organization of the SDS molecules and the existence of SDS aggregates adsorbed on the polymer chain, we carried out small-angle X-ray scattering experiments in solutions of PEO (0.100 wt %) in 1.00 and 10.0 mmol L$^{-1}$ Na$_2$[Fe(CN)$_5$NO] as well as solutions of PEO under the same conditions with 1.00 and 10.0 mmol L$^{-1}$ Na$_3$[Co(NO$_2$)$_6$]. In 10.0 mL flasks, we made additions of 75, 125, 175, 250, 350, and 775 µL of SDS (10.0 wt %) in each one of the four mother solutions. The corresponding SDS molar concentrations were 2.60, 4.33, 6.06, 8.66, 12.1, and 26.9 mmol L$^{-1}$, respectively.

All of the scattering curves corresponding to the solution intensity were subtracted from that of pure water. We observed that the signals due to the salts and to the PEO at the same concentration were negligible compared to water (results not shown here). In this way, we assumed that the scattered intensity was mainly due to the SDS molecules.

The scattering curves for PEO + SDS in the presence of Na$_2$[Fe(CN)$_5$NO] at two concentrations (1.00 and 10.0 mmol L$^{-1}$) are presented in Figure 7. In almost all curves, an oscillation in the scattered intensity was observed, which is characteristic of structures with a narrow size distribution.

The formation of SDS micelles was confirmed by the valley-and-peak shape of the scattering curves in the range of the scattering vector from 1 to 4 $nm^{-1}$, except for the most dilute solutions (75 and 125 μL for 1.00 mmol $L^{-1}$ and 75 μL for 10.0 mmol $L^{-1}$). These results suggested that either there were no micelles formed under those conditions or that the signal was not strong enough to show the oscillations. However, there was a $q$-dependence at low values of the scattering vector. One possible interpretation of this is that at low SDS concentrations, the molecules associate along the PEO molecules without self-organizing into micelles. This association would increase the macromolecule optical contrast, leading to an increase in the scattered intensity in the region of larger length scales (or small $q$ values).

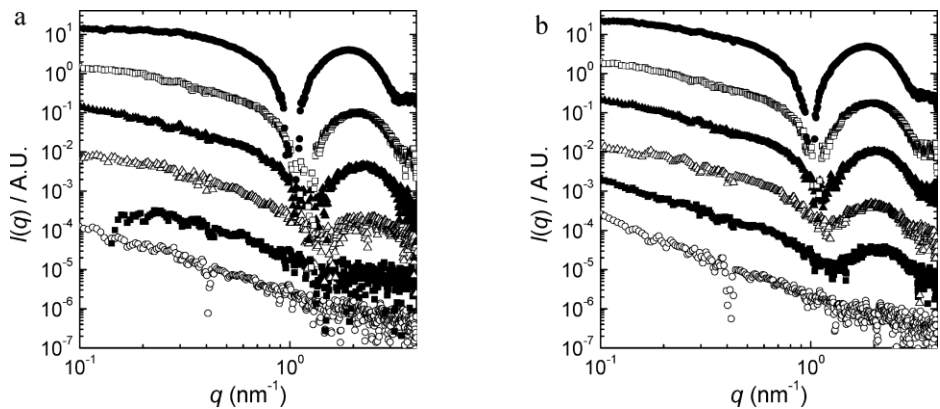

**Figure 7.** Scattered intensity of PEO35000 and SDS in $Na_2[Fe(CN)_5NO]$ aqueous solutions at (a) 1.0 and (b) 10.0 mmol $L^{-1}$. Each curve corresponds to a different volume of SDS solution of (○) 2.60, (■) 4.33, (△) 6.06, (▲) 8.66, (□) 12.1, and (●) 26.9 mmol $L^{-1}$.

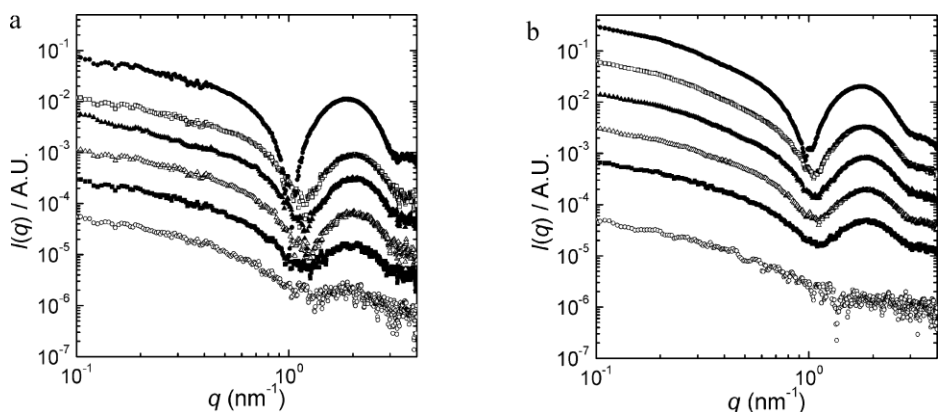

**Figure 8.** Scattered intensity of PEO35000 and SDS in $Na_3[Co(NO_2)_6]$ aqueous solutions at (a) 1.0 and (b) 10.0 mmol $L^{-1}$. Each curve corresponds to a different volume of SDS solution of (○) 2.60, (■) 4.33, (△) 6.06, (▲) 8.66, (□) 12.1, and (●) 26.9 mmol $L^{-1}$.

If we assume that micelles are spherical, then the scattering intensity is given by the well-known equation for monodisperse spherical structures[45]

$$I(q) \sim \frac{[\sin(qr) - qr\cos(qr)]^2}{(qr)^6} \quad (2)$$

where $r$ is the structure's radius. This function has a sinusoidal behavior with amplitude decreasing with increasing the $q$ values. It is straightforward to check that the first zero of the eq. 1 is located at $qr = 4.5$. The structures size polydispersity has the effect of damping the intensity oscillations, but the position of the minimum does not change significantly. So the micelles radius were calculated as $r = 4.5/q_{min}$ where $q_{min}$ is the position of the scattering vector corresponding to the first minimum of intensity. The micelles radius varied from 3.0 to 4.3 nm upon increasing the amount of SDS at salt concentrations of 1.00 mmol $L^{-1}$ and from 3.4 to 4.5 nm at 10.0 mmol $L^{-1}$, which agrees with previous results in the SDS + PEO systems.[46,47] One other effect that should be noted is that the intensity of the peaks at $q \approx 1.9$ $nm^{-1}$ increased with the SDS amount due to an increase in the amount of scattering (intensity is proportional to the SDS concentration). Finally, the form of the curves was not strongly altered by the salt concentration, meaning that 1.00 mmol $L^{-1}$ is sufficient for promoting micelle formation.

Analogous analyses were made using $Na_3[Co(NO_2)_6]$ at the same salt concentrations (1.00 and 10.0 mmol $L^{-1}$) with the same volumes of SDS solution (75-775 μL) (Figure 8). The results were analogous to those for $Na_2[Fe(CN)_5NO]$, but the intensity values were bigger than those observed using the previous salt, indicating that the formation of micelles is easier in the presence of $Na_3[Co(NO_2)_6]$ and that it is dependent on the salt concentration (the intensity increased when the salt concentration was changed from 1.00 to 10.0 mmol $L^{-1}$). It is important to point out that a small oscillation appeared in the curve with the smaller amounts of SDS and salt, indicating that it is possible that micelles are formed at all concentrations of the two salts, and that the intensity curves were not precise enough to show these processes.

The results of small-angle X-ray scattering corroborated the models proposed above. For the curve in a solution of PEO with $Na_2[Fe(CN)_5NO]$, the $\Delta H_{ap\text{-}int}$(dif) was reduced to almost zero, suggesting, possibly, a process in which there is no formation of aggregates. However, small-angle X-ray scattering showed that in the presence of $Na_2[Fe(CN)_5NO]$ and $Na_3[Co(NO_2)_6]$ aggregate formation did occur. The drastic changes in the isothermal titration calorimetric curves were attributed to changing interactions with the cosolute. Thus, the combination of these results leads us to conclude that the cosolute interacts with the molecules of PEO, competing with the surfactant for adsorption sites that exist along the polymer chain.

## 4. Conclusion

The present investigation revealed that the tendency of SDS molecules to interact with PEO does not change very much with the presence of simple ionic cosolutes. Enthalpic titration curves showed the same basic features in different systems. These profiles were significantly changed in the presence of the complex ionic cosolute $Na_2[Fe(CN)_5NO]$, which caused a large decrease in the apparent molar enthalpy change of interaction between PEO35000 and SDS ($\Delta H_{ap-int}(dif)$), suggesting the absence of polymer-surfactant interactions. However, the SAXS data showed that independent of the nature or concentration of ionic cosolutes, there were aggregates adsorbed on the polymer chain. The cosolutes NaCl, $Na_2SO_4$, $Li_2SO_4$, and NaSCN at concentrations of 10.0 and 100 mmol $L^{-1}$ caused a decrease in the cac values, indicating that association starts at lower surfactant concentrations, and promoted an increase in $C_2$. Both salt effects could be attributed to a decrease in the interfacial tension between the PEO surface and water and an increase in the surfactant chemical potential, respectively. The strong effect of the nitroprusside salt could be attributed to the specific interaction between the complex ionic solute and the EO unit of the polymer.

**Acknowledgment.**

We gratefully acknowledge Fundac¸a˜o de Amparo a Pesquisa do Estado de Minas Gerais (FAPEMIG), Conselho Nacional de Desenvolvimento Cientı´fico e Tecnolo´gico (CNPq), and Instituto Nacional de Cieˆncias e Tecnologias Analı´ticas Avanc¸adas (INCTAA) for financial support of this project. A.B.M. and G.M.D.F. thank CNPq, and I.J.B.S. thanks FAPEMIG for research fellowships. The National Laboratory of Synchrotron Light (LNLS, Campinas, Brazil) is acknowledged for the use of their facility.